# Parity-time symmetry in wavelength space with spatial singularity

**Authors:** Lingzhi Li[1], Guangying Wang[1], Jiejun Zhang*[1], Xinhuan Feng[1], Baiou Guan[1] and Jianping Yao*[1,2]

**Affiliations:**

[1]Guangdong Provincial Key Laboratory of Optical Fiber Sensing and Communications, Institute of Photonics Technology, Jinan University, Guangzhou 510632, China

[2]Microwave Photonics Research Laboratory, School of Electrical Engineering and Computer Science, University of Ottawa, Ottawa, ON K1N 6N5, Canada

*Correspondence to: zhangjiejun@jnu.edu.cn; jpyao@uottawa.ca

**Abstract:** Implementation of a parity-time (PT) symmetric microwave photonic system in the optical wavelength space with spatial singularity is proposed. In the proposed PT-symmetric microwave photonic system, the gain and loss modes are confined in a single spatial resonator, which is different from a conventional PT-symmetric system in which the two modes are localized in two physically separated resonators to form one-dimensional spatial potential symmetry as required by the simplest one-dimensional parity transformation. We show that PT-symmetry can be implemented between subspaces in non-spatial parameter spaces, in which the gain and loss modes can perfectly overlay spatially but are distinguishable in the designated parameter space. The resultant spatial singularity enables the possibility in implementing PT-symmetric systems with increased structural simplicity, integration density and long-term stability. To prove the concept, a PT-symmetric optoelectronic oscillator (OEO) in the optical wavelength space is implemented. The OEO has a single-loop architecture, with the gain and loss microwave modes carried by two optical wavelengths to form two mutually coupled wavelength-space resonators (WSRs). PT-symmetry is achieved by controlling the wavelength spacing and the power contrast. The operation of PT symmetry in the OEO is verified by the generation of a 10-GHz microwave signal with a low phase noise of 129.3 dBc/Hz at 10-kHz offset frequency and a high sidemode suppression ratio (SMSR) of 66.22 dB. Compared with a conventional spatial PT-symmetric system, one in the wavelength space features a much simpler configuration, better stability and greater resilience to environmental interferences.

**Main Text:**

Parity-time (PT) symmetric systems have attracted intensive research interests in the past few years (*1-3*). As a special non-Hermitian quantum system, a PT-symmetric system has real eigenvalues, which counters a common intuitive that real eigenvalues are only related to Hermitian observables (*4*). The discovery of PT symmetry has led to a burst of studies in both microscopic and macroscopic systems, including but not limited to atomic (*5*), electronic (*6-8*), thermal (*9*), photonic (*10-19*) and opto-electronic systems (*20, 21*). Photonic techniques have been proven to be a convenient approach to implement PT symmetry, where the complex refractive index distribution of a photonic system can be used to emulate the PT-symmetric potential such that $U(r)=U^*(-r)$, where * denotes complex conjugate and $r$ is the position vector (*1, 3*). So far, experimental implementations of PT symmetry are bounded to one-dimensional space where $r$ becomes a scalar, with most demonstrations focusing only on a PT-symmetric system with two subspaces (*6-11, 14-21*) and a few being extended to multiple subspaces (*12, 13*). One-dimensional PT symmetry formed by two subspaces is the simplest form of PT symmetry as the energy flow between the subspaces is the key to achieve the complementary gain and loss modes. In addition, the parity transformation in a simper system, e.g., a single subspace, becomes meaningless.

Since the advent of PT-symmetric photonics, a variety of optical devices and subsystems with novel functionalities emerge. Specifically, PT symmetry was employed for mode selection in a micro-ring laser by controlling the gain and loss interplay between two cavities (*14, 15*); PT symmetry was used to implement a coherent perfect absorber that can potentially realize extremely high modulation depth for future communications systems (*16, 17*); PT symmetry was used to implement optical non-reciprocity without the need for ferromagnetic materials and thus is suitable for large scale integration (*18, 19*). Recently, we demonstrated a parity-time symmetric opto-electronic oscillator (OEO) with two cross-coupled opto-electronic loops. By controlling gain and loss interactions, a single-frequency microwave signal with an ultra-low phase noise was generated (*20, 21*). This was the first time that PT symmetry was introduced to microwave photonic systems to overcome the long existing mode-selection challenge, which has severely jeopardized the development and wide applications of OEOs (*22-24*).

A significant common ground of all those PT-symmetric systems is that they consist of two spatially distributed subspaces, usually two resonators, to support the gain and loss supermodes. The successful realization of PT-symmetry relies critically on the perfect matching of the geometry of the two resonators and the availability of techniques to control the gain, loss and coupling between them. Compared to a Hermitian system, the redundant resonator in a PT-symmetric system makes it more precision-demanding in terms of fabrication or construction and more susceptible to interference especially for a macro system as the resonators may have vastly different localized perturbations. Hence, a simple question is if there is a technique to implement a PT-symmetric system with a single spatial resonator. Noting that such a PT-symmetric system has only one subspace. Energy flow between subspaces in a PT symmetric system is intuitively impossible, nor does the system has a degree of freedom to perform parity transformation.

Multiplexing is a well-established technique that is widely used in modern optical communications systems to increase the bandwidth, such as time-division multiplexing (TDM) and wavelength-division multiplexing (WDM). For TDM, an individual channel occupies a time slot and multiple channels are multiplexed in the optical traffic train. For WDM, an individual channel is modulated to one wavelength and multiple channels at multiple wavelengths are multiplexed and transmitted over an optical fiber (*25, 26*). TDM and WDM techniques are used to expand the capacity of a single optical fiber to that of multiple fibers, thus the spatial redundancy is reduced.

In this paper, we propose that PT symmetry can be implemented in a non-spatial parameter space, which is fundamentally different from any existing PT-symmetric system (*14-21*) by transferring the spatial duplicity of the PT-symmetric subspaces in to a non-spatial parameter space, and thus the system collapses into a zero-dimensional space with spatial singularity. As energy flow tunnels are properly built for power exchange between the gain and loss subspaces in the non-spatial parameter space, such a system can be stable and has an improved performance as compared with a spatially PT-symmetric system (*14-21*). A comparison between a spatial and a non-spatial (such as one in the wavelength space) PT symmetric system is shown in Fig. 1. For experimental verification, we design a microwave photonic system in a non-spatial parameter space of wavelength. The microwave photonic system supports microwave modes, which are modulated on two optical wavelengths at an optical modulator, transmit in a long fiber and recovered at a photodetector (PD). Such an hybrid configuration is generally known as an OEO that can be used for high quality microwave signal generation (*23*). Since two carrier wavelengths are employed, the system has two optoelectronic resonators in the parameter space of optical wavelength. The wavelength-space resonators (WSRs) are designed to be mutually coupled with independently controllable mode spacing and round-trip gains, but physically overlaps in space. We demonstrate that PT symmetry is achievable between the two WSRs, leading to single-mode oscillation of the OEO. The spatial simplicity is of great importance for practical applications of PT symmetry. The results show that the wavelength-space PT-symmetric OEO features the merits of higher stability, lower complexity and smaller footprint compared to its counterpart designed with the conventional methodology of a dual-spatial-loop arrangement (*20*). More importantly, the concept of multiplexing PT-symmetric subspaces in a non-spatial parameter space can be implemented using a variety of existing devices and subsystems. For photonic or microwave systems, several physical parameters can be chosen to form a parameter space for the implementation of PT symmetry, such as location, polarization, transverse mode, angular momentum and wavelength, if the parameter space is a Hilbert space that supports linear operation of eigenmodes. Specifically, in our experimental verification, a PT-symmetric OEO is implemented in the parameter space of optical wavelength.

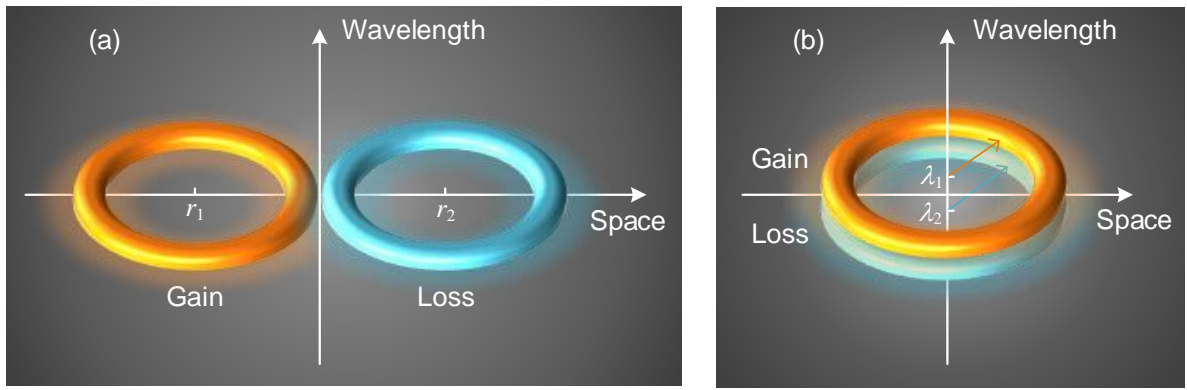

**Fig. 1.** Comparison of a spatial PT-symmetric system and a PT-symmetric system in the parameter space of wavelength. (a) A conventional PT-symmetric system with binary spatial arrangement with two subspaces located at $r_1$ and $r_2$, corresponding to our previous work in (*20*); (b) the proposed PT-symmetric system in the parameter space of wavelength with two subspaces located at $\lambda_1$ and $\lambda_2$. The two subspaces are non-distinguishable in terms of location, i.e., they fully overlap in space and the footprint is only half that of its spatial counterpart.

The schematic of the wavelength-space PT-symmetric system is shown in Fig. 2. It consists of two tunable laser sources (TLSs), a Mach-Zehnder modulator (MZM), a PD, an electric amplifier (EA), a long single-mode fiber (SMF), an electric bandpass filter (EBF) and a microwave power splitter. The two TLSs are used to generate two optical carriers with different wavelengths that are combined at a polarizer beam combiner (PBC) and sent to the MZM via a polarization controller (PC). The polarization directions of the optical carriers are orthogonal and are aligned with the principal axes the PBC to maximize the combination efficiency. Note that the MZM has a built-in polarizer, the tuning of the PC will change the polarization direction alignment between each carrier and the MZM, thus changes the carrier power injected into the OEO loop. In this way, the optical power ratio between the two optical carriers recirculating in the loop can be continuously tuned, while the total optical power applied to the MZM remains constant (see Supplementary Materials). The two optical carriers are modulated by oscillating microwave modes at the MZM, which then travels through a 10-km SMF and is then detected by the PD and converted back to electrical domain. After electrical amplification by the EA, the electrical signal is split into two parts by a 3-dB power splitter, with 50% of the output fed back to the MZM after the EBF. and the other 50% sent to an electrical spectrum analyzer for signal quality analysis. The combined operation of the PD, the EA, the EBF, the power splitter and the MZM corresponds to a microwave 2×2 coupler that allows the coupling between the microwave eigenmodes modulated to the two optical carriers (see Supplementary Materials).

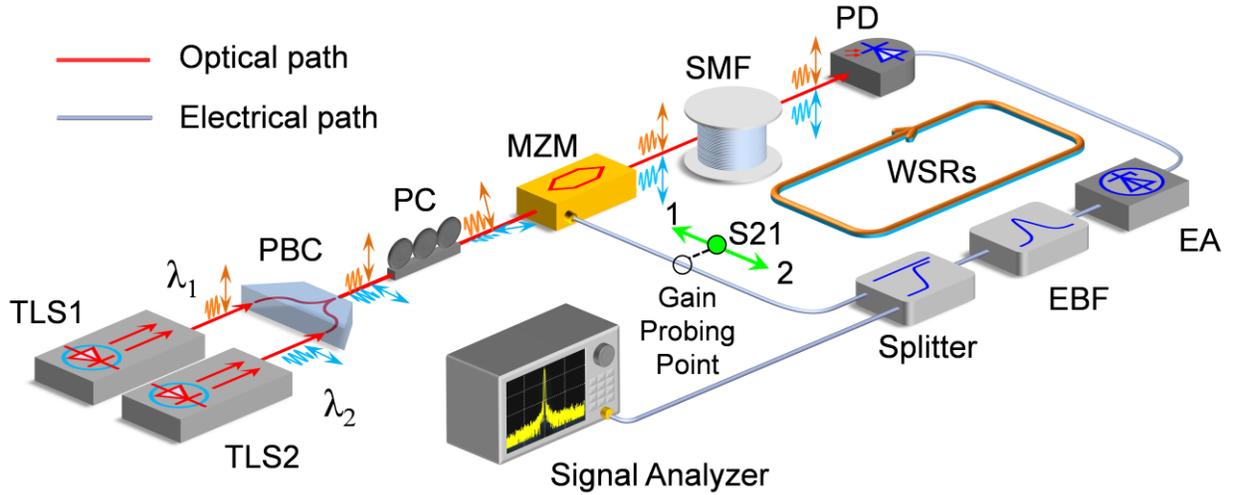

**Fig. 2.** Experimental implementation of a PT-symmetric OEO in the wavelength space. The system consists of a single optical and electrical hybrid spatial loop, in which two optical carriers of different wavelengths are used to allow the propagation of the microwave eigenmodes through the delay fiber. TLS: tunable laser source; PBC: polarizer beam combiner; PC: polarization controller; MZM: Mach-Zehnder modulator; SMF: single-mode fiber; PD: photodetector; EA: electric amplifier; EBF: electric bandpass filter; WSR: wavelength-space resonator.

Due to the chromatic dispersion of the SMF, the eigenfrequency of the local mode in a WSR is dependent on the optical carrier wavelength and is given by

$$\omega_m = \frac{2m\pi}{\tau_0 + D(\lambda - \lambda_0)} \qquad (1)$$

where $\omega_m$ is the angular frequency of the $m$-th order eigenmode; $\tau_0$ is the round trip time delay of a reference carrier wavelength of $\lambda_0$; $D$ is the dispersion coefficient of the SMF and $\lambda$ is the optical carrier wavelength from the TLSs. Since the two optical carriers are of different wavelengths, the PT symmetry across all modes of the two WSRs are not achievable. However, with the adoption of the narrowband EBF, PT symmetry can be achieved among the eigenmodes within the passband if the passband width of the electrical filter $\Delta f \ll 1/(\Delta\lambda \cdot D)$, and if the wavelength difference between the two carriers is (see Supplementary Material)

$$\Delta\lambda = \frac{n}{f_c \cdot D} \quad (2)$$

where $n$ is an integer and $f_c$ is the central passband frequency.

With PT symmetry, the coupling equations of the wavelength-space subsystems can be written as

$$i\frac{d}{dt}\begin{pmatrix} a \\ b \end{pmatrix} = \begin{pmatrix} \omega^{(1)} + i\gamma^{(1)} & -\kappa \\ -\kappa & \omega^{(2)} + i\gamma^{(2)} \end{pmatrix}\begin{pmatrix} a \\ b \end{pmatrix} \quad (3)$$

where $a$ and $b$ are the amplitudes of the localized eigenmodes in the two WSRs, of which the eigenvalues are $\omega^{(1)}$ and $\omega^{(2)}$; $\gamma^{(1)}$ and $\gamma^{(2)}$ are the gain or loss coefficients of the two WSRs and $\kappa$ is the coupling coefficient between them. With PT symmetry, we have $\omega^{(1)} = \omega^{(2)} = \omega_m$ and $\gamma^{(1)} = -\gamma^{(2)} = \gamma_m$ The eigenfrequencies of the supermodes in the PT-symmetric system can be calculated to be

$$\omega_m^{(1,2)} = \omega_m \pm \sqrt{\kappa^2 - \gamma_m^2} \quad (4)$$

PT-symmetry enhances the gain contrast between the modes with the highest and the second highest loop gain within the EBP bandwidth by a factor of $F = \sqrt{\gamma_m^2 - \gamma_n^2}/(\gamma_m - \gamma_n)$. It thus enables single-mode oscillation of the WSR-based PT-symmetric OEO, which would otherwise impossible since there are approximately 1000 modes within the EBP bandwidth.

Furthermore, the spatial singularity of a PT-symmetric OEO in the wavelength space also implies higher operation stability, as compared with a spatial PT-symmetric OEO under environmental disturbance such as temperature change or vibrations. Perturbation imposed on a PT-symmetric system can be decomposed into common mode and differential mode signals with respect to the two subsystems. For WSR-based PT-symmetric OEO, the temperature or vibration disturbance affects two subsystems indiscriminately, the perturbed Hamiltonian of the system is thus given by

$$H = \begin{pmatrix} \omega_m + i\gamma_m + \varepsilon_{\omega_m^{(1)}} & -\kappa \\ -\kappa & \omega_m - i\gamma_m + \varepsilon_{\omega_m^{(1)}} \end{pmatrix} \quad (5)$$

where $\varepsilon_{\omega_m^{(1)}}$ is the localized eigenfrequency perturbation induced by temperature change or vibrational interferences (27) (see Supplementary Materials). The corresponding perturbed eigenfrequency of the supermodes is $\omega_m = -\varepsilon_{\omega_m^{(1)}}$, indicating that in the presence of environmental interferences, no eigenfrequency bifurcation will appear to affect the stability of PT symmetry. Instead, the eigenfrequency of both the gain and loss supermodes will drift at the same rate, ensuring that PT-symmetry is always preserved. Though temperature can also affect the dispersion coefficient of the SMF and thus affect the PT symmetry, it is proven that such effect is small and negligible within a practical application scenario.

## Discussion

An experiment is carried out based on the setup shown in Fig. 1. We first measure the open-loop frequency response of the OEO loop at the gain probing point by opening the feedback loop and adding a vector network analyzer (VNA) between the MZM and the splitter. The electrical amplifier ensures the open-loop response of the OEO is greater than unity at certain frequency band, so that modes within the frequency band can exceed the oscillation threshold. Fig. 3(a) shows the measured open-loop frequency response of the OEO loop, in which the EBP with a central frequency of 10 GHz and a bandwidth of 20 MHz is deployed. The frequency response has a smooth profile, as shown in Fig. 3(b). Without PT symmetry, it is estimated that there are about 1000 longitudinal modes within the passband. Single-mode oscillation would not be possible without the adoption of PT symmetry.

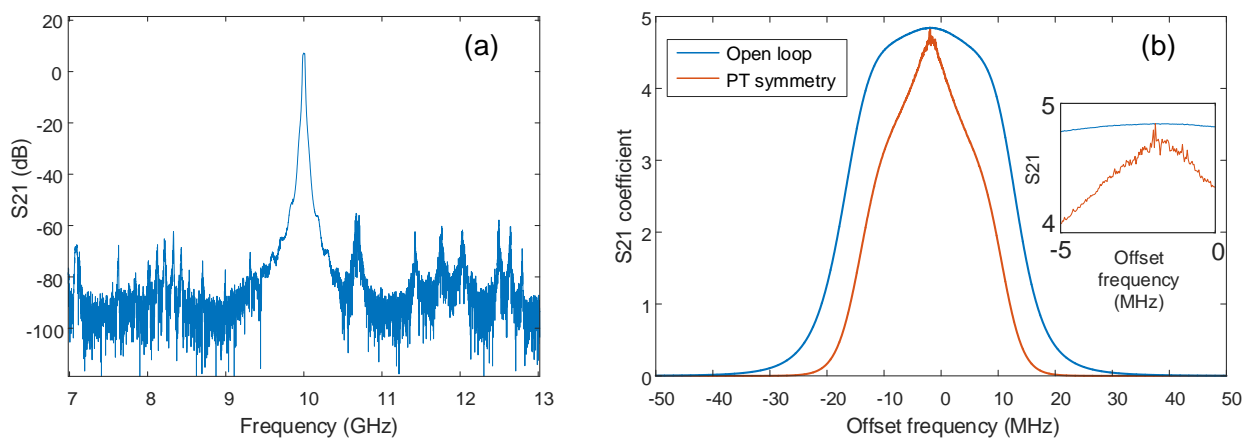

**Fig. 3.** The open-loop gain spectrum. (a) Gain spectrum measurement by a VNA connected at the gain probing point; (b) Zoom-in view of the gain spectrum at 10 GHz (blue) and the calculated gain profile (red) after gain contrast enhancement with PT symmetry. Inset: zoom-in view of the enhanced ripples.

Then, we close the OEO loop and oscillation will start. We analyze the generated microwave signal using an electrical spectrum analyzer (ESA). If only one TLS is turned on, the system is a single-loop OEO without PT symmetry. Multimode oscillation is achieved, as shown in Fig. 4(a). To achieve single-mode oscillation, the two TLSs are turned on with a wavelength spacing calculated by Eq. (2). The light waves from the TLSs are orthogonally polarized and are combined at the PBC. Thanks to the built-in polarizer in the MZM, the tuning of the PC can result in a change in the power ratio between the two wavelengths in the feedback loop, i.e., by tuning the PC, the WSR corresponding to one wavelength can have a round-trip gain, while the WSR corresponding to the other wavelength can have a round-trip loss. In this way, PT symmetry is achieved between the two WSRs. The small gain ripples on the open-loop gain spectrum is enhanced to large ripples, as shown in Fig. 3(b), and the single-mode oscillation is achieved. The spectrum of the generated signal with a 100-MHz measurement span in Fig. 4(b) shows that a single longitudinal mode oscillation is achieved near the frequency of 10 GHz. Fig. 4(c) and (d) shows the spectrum with two different spans of 100 kHz and 1 kHz. The longitudinal mode spacing of the OEO is about 20 kHz and the oscillating mode is 46.75 dB higher over the highest side mode. Thanks to the single spatial loop architecture, we observe that the OEO operates stably without mode hopping within one hour of monitoring time without any isolation from the interference in a laboratory environment. Our analysis shows that the PT-symmetric system in wavelength space can be over 1000 times more resilient to environmental perturbation as compared with its spatial counterpart (see Supplementary Materials).

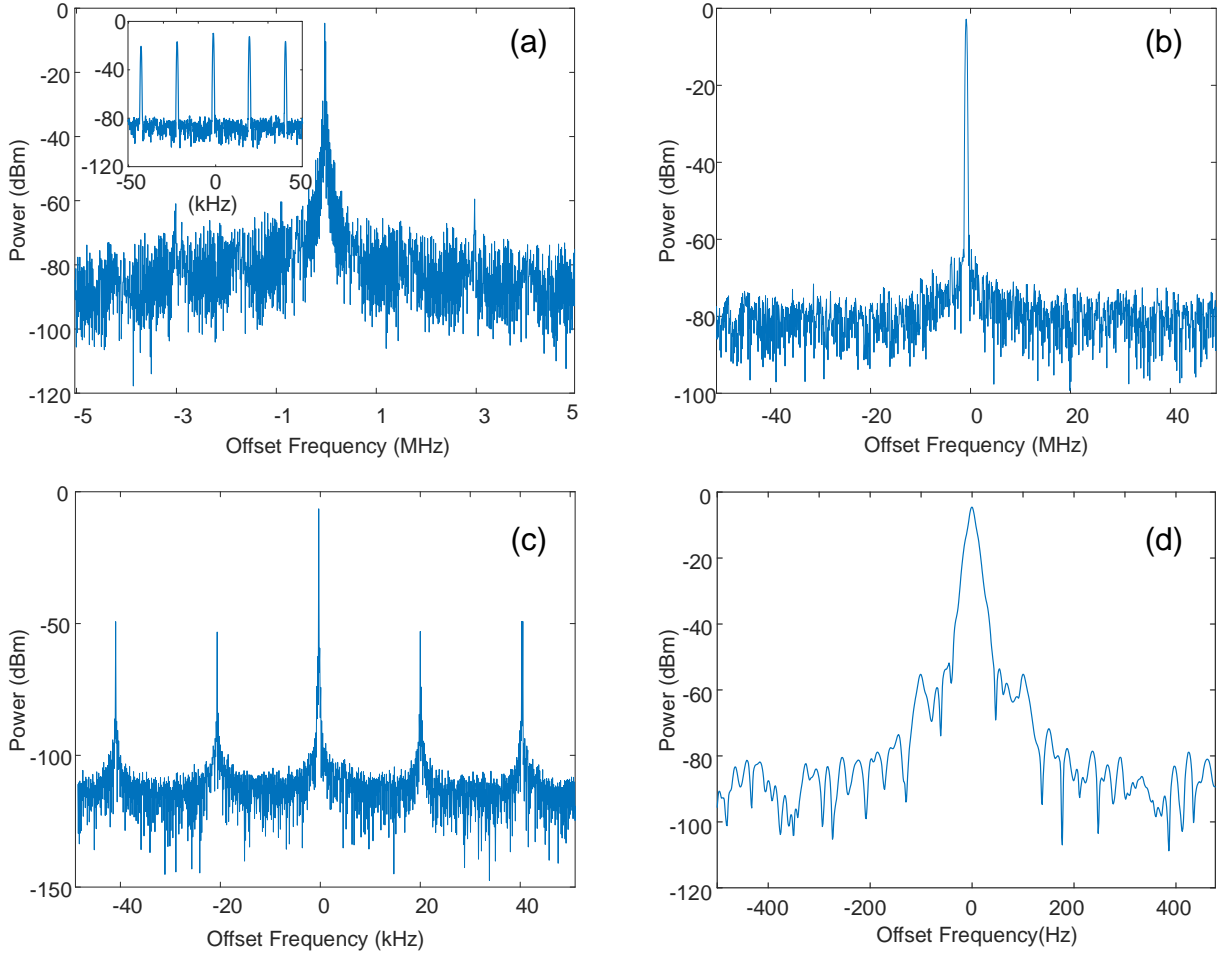

**Fig. 4.** The electrical spectra of the microwave signals generated by the wavelength-space PT-symmetric OEO. The spectra are measured at a central frequency at 10 GHz. (a) Multimode oscillation spectrum measured with a resolution bandwidth (RBW) of 3 MHz. Inset: zoom-in view of the multimode spectrum showing multiple modes with comparable amplitudes; single-mode oscillation spectra measured with RBWs of (b) 3 MHz, (c) 100 kHz and (d) 9 Hz. The spectrum in (c) shows a dominating mode with a sidemode suppression ration of 46.75 dB.

Figure 5 shows the measured phase noise of the generate microwave signal. The phase noise is -129.3 dBc/Hz at an offset frequency of 10 kHz with sidemodes lower than -66.22 dBc/Hz, which verifies the effectiveness of the mode-selection mechanism in the WSR-based PT-symmetric OEO. For comparison, the phase noises of two signals generated by a commercial microwave signal generator (Agilent E8254A) and a spatially dual-loop PT-symmetric OEO are also shown in Fig. 5. At an offset frequency of 10 kHz, the phase noise of the WSR-based PT-symmetric

OEO is 13.4 dB lower than that of the commercial microwave source, but 13.0 dB higher than that of a dual-spatial loop OEO due to the higher noise floor of the measurement instrument (Keysight E9040B signal analyzer) that we used in the experiment. Since the phase noise performance of an OEO is mainly determined by the length of the SMF within the loop, we estimate that the actual phase noise of the OEO is around -140 dBc/Hz at 10 kHz offset frequency if a measurement instrument with a sufficiently low noise floor is used (*22, 23*).

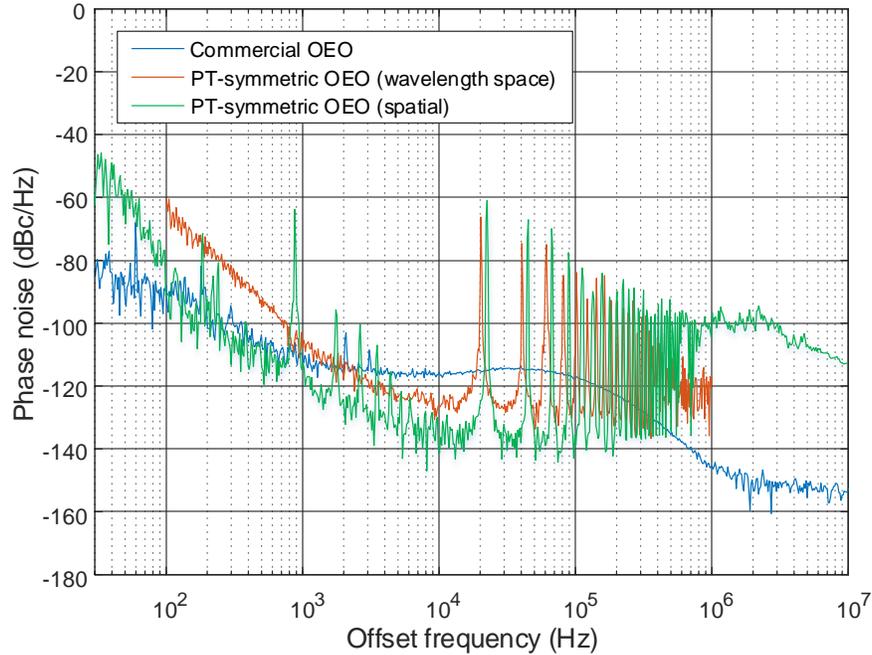

**Fig. 5.** Measured phase noise of the generated microwave signal (red). For comparison, the phase noises of microwave signals generated by a commercial microwave source (blue) and a spatially dual-loop OEO with a 9.1-km loop length (green) are also shown.

In conclusion, we have demonstrated that PT symmetry can be implemented in the parameter space of optical wavelength, to eliminate the requirement for physically separated spatial subsystems in a conventional PT-symmetric system. An OEO with two optical carriers was implemented to validate the concept. Single-mode oscillation was realized with the single-spatial-loop OEO, in which a low phase noise microwave signal was generated with a high sidemode suppression ratio. The WSR-based PT-symmetric OEO can be a new paradigm for high performance microwave sources as it is over 1000 times more resilient to environment interference, and it is much easier to implement due to the structural simplicity and compactness compared to its conventional spatial counterpart (*20*). The wavelength-space PT-symmetric OEO features spatial singularity within a single closed loop, which counters common knowledge that a PT-symmetric system must be an open system that interacts with the environment.

The concept of constructing PT-symmetric systems in physical parameter spaces other than those only based on spatial arrangements has enabled the possibility toward a significantly enhanced performance and functionality. For example, in an integrated PT-symmetric system, spatial compactness due to the implementation in a non-spatial parameter space can increase the operation stability, double the integration density and thus reduce the cost; the multi-dimensional parameter space of location, polarization, transverse mode, angular momentum and wavelength can be used to implement complex PT-symmetric network, which enables the controllable multi-dimensional energy flow and may find great applications in future neuro-networks.

**Acknowledgments:** We acknowledge Keysight Technologies and Rohde & Schwarz for providing instruments for signal quality analysis.

**Funding:** This work is supported by the National Natural Science Foundation of China (61905095) and Tier 3 Talent Recruitment Program of Jinan University.

**Author contributions:** J. Z. conceived the idea, designed the experiment and wrote the manuscript; L. L. and G. W. performed the experiment; J. Y supervised the project and revised the manuscript; X. F. and B. G. provided resources.

**Competing interests:** Authors declare no competing interests.

**Data and materials availability:** All data is available in the main text or the supplementary materials.

**Supplementary Materials:**

Materials and Methods

Figures S1-S4

Movie S1

References (*1-8*)

# Supplementary Materials for

Parity-time symmetry in wavelength space with spatial singularity

Lingzhi Li[1], Guangying Wang[1], Jiejun Zhang[1]*, Xinhuan Feng[1], Baiou Guan[1] and Jianping Yao[1,2]

Correspondence to: zhangjiejun@jnu.edu.cn and jpyao@uottawa.ca

**This PDF file includes:**

Supplementary Text

Figs. S1 to S4

Caption for Movie S1

**Other Supplementary Materials for this manuscript include the following:**

Movie S1

**Supplementary Text**

1. Wavelength-space resonators and eigenmodes

In an optical and electrical hybrid loop or resonator, the round-trip time is a summation of the time delays introduced by the electrical and optical paths. The time delay of the electrical path is much smaller than that of the optical path, and is independent of the optical carrier wavelength, hence it is ignored in our analysis. The time delay of the optical path is mainly due to the long optical fiber which is usually a single mode fiber (SMF). The total round-trip time delay can be approximately expressed as

$$\tau(\lambda) = \tau_0 + D \cdot (\lambda - \lambda_0) \quad \text{(S1)}$$

where $\lambda$ is the carrier wavelength; $\tau_0$ is the time delay of a reference wavelength $\lambda_0$ as it travels through the SMF and $D$ is the dispersion coefficient of the SMF. Note that the hybrid resonator does not contain a closed loop for the optical carrier. At the photodetector, the microwave signal modulated on the optical carrier will be detected and is directed to the MZM via the RF input port to form a closed microwave loop. Hence, the optical carrier only plays a role to carry the microwave signal to enable its transmission in the optical fiber and the value of wavelength can be controlled to adjust the round-trip time, or the eigen frequency of the resonator in the microwave regime. The free spectral range (FSR) of the resonator is given by

$$FSR(\lambda) = \frac{1}{\tau(\lambda)} \quad \text{(S2)}$$

If two wavelengths are employed, two sets of eigenfrequencies will be achieved due to chromatic dispersion of the fiber, which corresponds to two sets of localized eigenfrequencies of two resonators in the wavelength space. Perfect match of all eigenfrequencies between the two wavelength space resonators (WSRs, denoted as WSR1 and WSR2 for the equivalent gain and loss resonators, respectively) cannot be achieved due to the different FSRs. We use a microwave bandpass filter in the electrical path, which has as a central frequency of $f_c$ and a bandwidth of $f_{BW}$ to ensure the symmetry of real part of the eigenfrequencies within the passband. As illustrated in Fig. S1, the FSR mismatch within the passband should be small and negligible if

$$\left| \frac{f_{BW}}{FSR(\lambda_1)} - \frac{f_{BW}}{FSR(\lambda_2)} \right| \ll 1 \quad \text{(S3)}$$

or

$$D \cdot f_{BW} \cdot \Delta\lambda \ll 1 \quad \text{(S4)}$$

where $\Delta\lambda = |\lambda_1 - \lambda_2|$ is the wavelength difference between the two optical carriers and $f_{BW}$ is the bandwidth of the electrical filter. Such a condition can be easily satisfied by using two wavelengths with a small wavelength separation and a narrowband filter, as shown in Fig. S1.

Although Eq. (S4) shows that small wavelength spacing between the optical carriers is more preferable for the WSRs to have a negligible FSR difference, the exact carrier-wavelength spacing should be properly chosen according to the dispersion coefficient of the optical fiber, so that the eigenfrequencies of one of the WSRs can be aligned to those of the other within the microwave filter passband. To quantify carrier-wavelength spacing, we consider that the system oscillates at the central frequency of the electrical bandpass filter $f_c$, which is also the frequency of the two degenerate eigenmodes $\omega_m^{(1)}$ and $\omega_n^{(2)}$, i.e., the $m$-th and $n$-th order modes of WSR1 and WSR2, respectively. The two eigenmodes can be expressed as

$$\omega_m^{(1)} = \frac{2m\pi}{\tau(\lambda_1)},$$
$$\omega_n^{(2)} = \frac{2n\pi}{\tau(\lambda_2)} \tag{S5}$$

By letting $\omega_m^{(1)} = \omega_n^{(2)}$, we get

$$\Delta\lambda = \frac{m-n}{D \cdot f_c} \tag{S6}$$

Since $(m-n)$ is an integer, valid wavelength spacings are a set of discrete values that are inversely proportional to the product of the oscillation frequency and fiber dispersion coefficient.

In our experiment, the two WSRs are used to form an oscillator that support single-mode oscillation at a frequency of 10 GHz. The dispersion coefficient of the long fiber (10-km SMF) is 170 ps/nm. The wavelength spacing is 2.353 nm, where we choose $(m-n) = 4$. Such a wavelength spacing can be implemented by using high accuracy tunable laser sources (TLSs). Moreover, the bandwidth of the electrical filter is $f_{BW} = 20 \, \text{MHz}$, substitute this value to Eq. (S4), we have $D \cdot f_{BW} \cdot \Delta\lambda = 0.008 \ll 1$. Such an alignment accuracy is higher than those reported in (*1-3*), which indicates that parity-time (PT) symmetry can be achieved across all modes within the electrical filter passband.

2. Tuning of round-trip gain in a WSR

In addition to the alignment of eigenfrequency, which is related to the real part of the potential function in a PT-symmetric system, the gain and loss tuning the WSR is also critical as it is related to the symmetry of the imaginary part of the potential. A photonic or microwave PT-symmetric system requires that the gain and loss coefficients are of the same magnitude in the two subsystems, i.e., the gain coefficient of one of the WSRs is equal to the loss coefficient of the other. In our system, the gain and loss coefficient adjustment is achieved by changing the powers of the two optical carriers.

In a microwave path, the gain or loss coefficient is also known as the S21 parameter, which is measured by the ratio between the microwave power returned from the path and that injected into the path. Here, we choose a probing point for the round-trip gain between the MZM and the microwave splitter, which is experimentally measured by a microwave network analyzer, as shown in Fig. S2.

First, we assume the MZM is biased at the quadrature point and an electrical signal given by $e(t) = V_{in}\cos(\omega t)$ is applied to the MZM. with an amplitude of $V_{in}$. The electrical field of the electrical signal can be given as $e(t) = V_{in}\cos(\omega t)$; where $V_{in}$ and $\omega$ are the amplitude and angular frequency of the electrical signal, respectively. Then, the optical power at the output of the MZM is given by

$$P_{out}^{(o)} = \frac{1}{2}\alpha_{MZM} P_{in}^{(o)} \left[1 - \sin\left(\frac{\pi}{V_\pi} e(t)\right)\right] \tag{S7}$$

where $P_{in}^{(o)}$ is the power of the optical carrier that is launched into the MZM; $\alpha_{MZM}$ and $V_\pi$ are the insertion loss and the half-wave voltage of the MZM, respectively. The modulated signal then travels through the SMF and is detected at the photodetector (PD). The photocurrent at the output of the PD is given by

$$\begin{aligned} i_{PD}(t) &= \alpha_{SMF} \Re P_{out}^{(o)} \\ &= \frac{1}{2}\alpha_o \Re P_{in}^{(o)} \left[1 - \sin\left(\frac{\pi}{V_\pi} e(t)\right)\right] \end{aligned} \tag{S8}$$

where $\Re$ is the responsivity of the PD; $\alpha_{SMF}$ is the insertion loss of the SMF and $\alpha_o = \alpha_{MZM}\alpha_{SMF}$ is the total optical loss. The photocurrent from the PD is then converted to a voltage signal by a transimpedance electrical amplifier (EA),

$$\begin{aligned} V_{PD}(t) &= R_{EA} i_{PD}(t) \\ &= -\alpha_o R_{EA} \Re P_{in}^{(o)} J_1\left(\frac{\pi V_{in}}{V_\pi}\right)\cos(\omega t) \end{aligned} \tag{S9}$$

where $R_{EA}$ is the load resistance of the EA and $J_1$ is the first order Bessel function of the first kind. It should be noted that in Eq. (S9), we apply Jacobi-Anger expansion to the amplitude-modulated signal and ignore all DC and higher order harmonic components, which is the case in our proposed system due to the existence of a DC block in the EA and the incorporation of the EBF to filter out undesirable frequencies. The signal becomes a single-tone signal with a frequency equal to that of the input signal at the probing point. We can calculate the electrical power, given by

$$P_{PD}^{(e)} = \frac{\tilde{V}_{PD}^2}{2R_{EA}} = \frac{1}{2}\alpha_o^2 R_{EA} \Re^2 P_{in}^{(o)^2} J_1^2\left(\frac{\pi V_{in}}{V_\pi}\right) \tag{S10}$$

where $\tilde{V}_{PD}$ is the amplitude of the electrical signal that is sent to the EA. The signal is then amplified by the EA, experiences losses from the splitter and the EBF and finally reaches the probing point. The gain or loss coefficient is given by

$$G_{loop} = \frac{\alpha_e G_A P_{PD}^{(e)}}{V_{in}^2/(2R_{MZM})} = \alpha_e G_A \alpha_o^2 R_{EA}^2 \Re^2 P_{in}^{(o)^2} \cdot J_1^2\left(\pi V_{in}/V_\pi\right)/V_{in}^2 \tag{S11}$$

where $\alpha_e$ is the total electrical insertion loss contributed by the splitter and the EBF and $G_A$ is the gain factor of the EA. We assume that the load resistance of the MZM is the same as that of the EA. It can be seen from Eq. (S11) that the round trip gain is proportional to the gain and loss factors contributed by the electrical path, and to the square of those contributed by the optical path. The term $J_1^2\left(\pi V_{in}/V_\pi\right)/V_{in}^2$ represents a saturation effect induced by the nonlinearity of the electro-optic intensity modulation (*4*) that is only prominent when the amplitude of the signal is comparable to the half-wave voltage $V_\pi$. To allow the gain and loss to be tunable, we observe that the gain or loss coefficient of the loop is related to the square of the power of the optical carrier. When two optical carriers are used, the loop gain of each subsystem can be tuned by changing the optical power of that individual carrier.

The essence of implementing PT symmetry is to adjust the gain difference between two WSRs such that when the system reaches equilibrium, one loop will have a net gain and the other will have a net loss. According to the discussion above, this can be implemented by changing the optical power ratio between the two carriers. We use a polarization beam combination scheme to adjust such power ratio, while the total power of the two carriers are maintained nearly constant. This will guarantee that the PD works at its optimum input power level (*4*).

The principle of the polarization beam combination scheme is illustrated in Fig. S3. The two linearly polarized optical carriers of different wavelengths are generated by two tunable laser sources (TLSs) with equal power of $P_0^{(o)}$ and orthogonal polarization, which are combined at a polarization beam combiner (PBC). The polarization controller (PC) following the PBC is configured such that it does not vary the linear polarizations of the optical carriers, but only rotates the polarization directions by an angle of $\theta$. This can be achieved by ensuring that the optical axes of the equivalent quarter wave plates are aligned to the polarization directions of the optical carriers, and that the optical axes of the equivalent half wave plate intersect with the polarization directions of the optical carriers with an angle of $\theta/2$. We assume that the direction of the polarizer integrated in the MZM assembly is aligned with carrier $\lambda_1$ before rotation. The optical power of the two carriers travelling through the PC, the polarizer and finally reaching the MZM assembly is given by

$$P_{\lambda_1} = P_0^{(o)} \cos^2(\theta) \tag{S12}$$

$$P_{\lambda_2} = P_0^{(o)} \sin^2(\theta) \tag{S13}$$

substituting to (S11), the gain and loss coefficients of the two WSRs $G_{\lambda_1}$ and $G_{\lambda_2}$ become

$$G_{\lambda_1} = G_{max} \cdot \cos^2(\theta) \tag{S14}$$

$$G_{\lambda_2} = G_{max} \cdot \sin^2(\theta) \tag{S15}$$

$$G_{max} = \alpha_e G_A \alpha_o^2 R_{EA}^2 \Re^2 P_0^{(o)^2} \cdot J_1^2(\pi V_{in}/V_\pi)/V_{in}^2 \tag{S16}$$

where $G_{max}$ is the maximum round-trip gain of a WSR, which is achieved when the carrier polarization is perfectly aligned to that of the polarizer. For PT symmetry, it is required that the two WSRs have gain and loss coefficients of the same magnitude, i.e., $G_{\lambda_1} \cdot G_{\lambda_2} = 1$, we solve that

$$\theta = \pm \arcsin(\frac{2}{G_{max}}) \tag{S17}$$

To ensure that a solution exists for $\theta$, $G_{max}$ must be greater than 2. The two solutions indicate that the polarization rotation angle of $\theta$ can be performed on either optical carrier, and the other will be rotated by $\theta + 180°$, which leads to the identical PT-symmetry condition.

3. Coupling between WSRs

In addition to the match of eigenfrequencies, the gain and loss coefficients, the coupling between the WSRs are also mandatory for a PT-symmetric system (*2*).

In the proposed wavelength-space PT-symmetric optoelectronic oscillator (OEO), electrical signals in the two WSRs are launched into the PD simultaneously. Signal amplitude addition of the two signals in the two WSRs is automatically achieved at the output of the PD. The combined signal then propagates through the electrical path that is shared by the two WSRs and is applied to the MZM. It can be seen from Eq. (S7) that both optical carriers are modulated with the same modulation depth $P_{in}^{(o)}/P_{out}^{(o)}$, i.e., the signal is distributed into the two WSRs. From a microwave perspective, signals are combined and distributed indistinctively from and to the WSRs. Thus, the operation of the electrical path is equivalent to a microwave coupler with a coupling ratio of 1:1. Compared with a regular passive coupler, the incorporation of the EA in the electrical path can result in a negative insertion loss due to signal amplification. In the equivalent coupler, the powers of the optical carriers do not affect the coupling ratio, but the round-trip gain of the respective WSR.

4. PT symmetry based on two WSRs

A PT-symmetric system consists of two coupled cavities. The coupling equation of the system is given by

$$i\frac{d}{dt}\begin{pmatrix} a \\ b \end{pmatrix} = \begin{pmatrix} \omega^{(1)} + i\gamma^{(1)} & -\kappa \\ -\kappa & \omega^{(2)} + i\gamma^{(2)} \end{pmatrix}\begin{pmatrix} a \\ b \end{pmatrix} \qquad (S18)$$

where $\omega^{(1)}$ and $\omega^{(2)}$ are the eigenfrequencies of the two WSRs without gain, loss or coupling; $a$ and $b$ are the amplitude of the eigenmodes in the PT-symmetric system; $\gamma^{(1)}$ and $\gamma^{(2)}$ are the gain or loss coefficients of the two WSRs and $\kappa$ is the coupling between the two WSRs. In a PT-symmetric system, it is required that $\omega^{(1)} = \omega^{(2)} = \omega_m$ and $\gamma^{(1)} = -\gamma^{(2)} = \gamma_m$, where $m$ denotes the order of the primary eigenmode that is within the passband of the EBF. The eigenfrequencies of the PT-symmetric system can be given by

$$\omega_m^{(1,2)} = \omega_m \pm \sqrt{\kappa^2 - \gamma_m^2} \qquad (S19)$$

Considering a primary mode ($m$-th order) and a secondary mode ($n$-th order) with net loop gain coefficients of $\gamma_m$ and $\gamma_n$, which are respectively the modes with the highest and the second highest net loop gain among all modes, the gain different between the $m$-th and $n$-th order modes with PT symmetry can be calculated by (*2*)

$$\Delta g_{PT} = \sqrt{\gamma_m^2 - \gamma_n^2} \qquad (S20)$$

Note that a single WSR system without PT symmetry is given by $\Delta g = \gamma_m - \gamma_n$. The WSR-based PT-symmetric system provides an enhanced gain difference between the primary and the secondary modes, thus single-mode oscillation can be easily achieved simply by using the enhanced ripples on the frequency response of the WSRs. The single-mode oscillation can be used to verify the operation of PT symmetry in the wavelength space.

5. Carrier wavelength perturbation

In the proposed architecture, the two WSRs are formed by two optical carriers generated by two TLSs. As discussed in the previous Section, the two sets of eigenmodes corresponding to the two WSRs are the essence of the coupled subsystems required in a PT-symmetric system. In WSRs, the eigenmodes are tuned by tuning the wavelengths of the optical carriers, in contrast to those of the spatial PT-symmetric systems where the eigenmodes are tuned by the lengths of the spatial loops. Hence, these two kinds of PT-symmetric systems have different behaviors in case of environmental perturbations, including carrier wavelength drift, loop length change and temperature change. In this part, we investigate the effect of optical carrier wavelength drift on the operation of the PT-symmetric system.

First, we can assume that only one of the carrier wavelengths is perturbed, which is given by $\lambda_1 + \varepsilon_\lambda$. It can be derived from Eqs. (S1) and (S5) that the carrier wavelength perturbation will be transferred into the eigenvalue perturbation of the corresponding WSR, which is

$$\varepsilon_{\omega_m^{(1)}} = -\frac{2m\pi D}{\left[\tau_0 + D\cdot(\lambda_1 - \lambda_0)\right]^2}\varepsilon_\lambda \tag{S21}$$

In our experiment, a 10-km standard ITU G.652 SMF is used, and the oscillating microwave signal has a frequency of 10 GHz. The approximate values for the quantities in (S21) are $m = 4.8\times10^5$, $D = 170$ ps/nm and $\tau_0 + D\cdot(\lambda^{(1)} - \lambda_0) = 4.8\times10^{-4}$ s. Substitute to Eq. (S21), we have

$$\varepsilon_{\omega_m^{(1)}} = 2.2\times10^3\left(\text{rad/s}\times\text{nm}^{-1}\right)\cdot\varepsilon_\lambda \tag{S22}$$

which indicates that for 1 nm of carrier wavelength drift, the eigenfrequency of the corresponding WSR will drift for 350 Hz (converted from angular frequency). The wavelength stability of the TLSs used in the experiment (Keysight N7714A) are 2.5 pm in 24 hours, indicating that the eigenfrequency drift will be less than 0.88 Hz, which is significantly less than the FSR of the WSRs or the oscillating frequency of the PT-symmetric system. Hence, optical carrier wavelength drift has small and negligible effect on the stable operation of the PT symmetric system. The WSR-based PT-symmetric OEO can operate with great stability.

For a PT-symmetric system in a critical state that $\gamma_m = \kappa$, the perturbed Hamiltonian can be written as

$$H = \begin{pmatrix} \omega_m + i\gamma_m + \varepsilon_{\omega_m^{(1)}} & -\kappa \\ -\kappa & \omega_m - i\gamma_m \end{pmatrix} \tag{S23}$$

Letting $\det(H) = 0$, we have

$$\omega_m = \frac{-\varepsilon_{\omega_m^{(1)}} \pm \sqrt{\varepsilon_{\omega_m^{(1)}}^2 + 4i\kappa\varepsilon_{\omega_m^{(1)}}}}{2} \tag{S24}$$

Apparently, the wavelength perturbation on only one of the carrier wavelengths will increase the eigenmode mismatch between the two WSRs, induce an eigenvalue bifurcation of $\sqrt{\varepsilon_{\omega_m^{(1)}}^2 + 4i\kappa\varepsilon_{\omega_m^{(1)}}}$ and thus affects the stability of the PT symmetry operation of the system. Thus, it is important to use TLSs with good wavelength stability to ensure a small $\varepsilon_{\omega_m^{(1)}}$.

On the other hand, if both carrier wavelengths are perturbed for the same amount, the perturbed Hamiltonian will be

$$H = \begin{pmatrix} \omega_m + i\gamma_m + \varepsilon_{\omega_m^{(1)}} & -\kappa \\ -\kappa & \omega_m - i\gamma_m + \varepsilon_{\omega_m^{(1)}} \end{pmatrix} \tag{S25}$$

Letting $\det(H) = 0$ gives

$$\omega_m = -\varepsilon_{\omega_m^{(1)}} \tag{S26}$$

The unique solution indicates that the no eigenvalue bifurcation will take place in this situation. The PT symmetry will be preserved with good stability as the eigenvalues of the two WSRs are shifting for the same amount toward the same direction, ensuring that the PT symmetry condition is always met.

Compare Eq. (S23) with Eq. (S25), the difference is that the perturbation is applied to one WSR for Eq. (S23) and to both WSRs for Eq. (S25). The combination of the two kinds of perturbation can represent any realistic perturbation that a PT-symmetric system may experience, as a perturbation can always be decomposed to common mode perturbation and differential mode perturbation.

6. Stability under ambient temperature and vibration perturbation

For low phase noise microwave signal generation, a PT-symmetric OEO is usually required to have a long loop length, which will result in a low phase noise. A dual-spatial loop PT-symmetric system can effectively solve the mode-selection problem for a long loop that has a small FSR (*2*). However, the dual-spatial loop architecture also makes the system more susceptible to ambient temperature or vibration interferences, which will cause eigenmode mismatch between the two subsystems. In this part, we investigate the response of the wavelength-space PT-symmetric oscillator to the environment interference and verify that a wavelength-space PT-symmetric system has better stability compare to a spatial PT-symmetric system.

The ambient temperature and vibration interferences induce time-varying optical lengths for both WSRs, which arises due to physical elongation of fiber and photoelastic effect for vibration or strain interferences, and due to thermal expansion of fiber and the temperature-dependent refractive index for temperature interferences. The combined effects are the change of the optical length of the SMF. Without losing generality, we can assume that the length of the optical fiber remains constant and attributes all effects induced by temperature change and vibration to the change of effective refractive index of the SMF. Thus, the interference-induced refractive change can be written as (*5*):

$$\frac{\varepsilon_{n_{eff}}}{n_{eff}} = \left[1 - \frac{n_{eff}^2}{2}\left(P_{12} - \nu P_{11} - \nu P_{12}\right)\right]\varepsilon_s + \left[\alpha + \frac{1}{n_{eff}}\frac{dn_{eff}}{dT}\right]\varepsilon_T \tag{S27}$$

where $\varepsilon_s$ and $\varepsilon_T$ denotes the magnitudes of strain and temperature disturbance, respectively; $P_{i,j}$ are the Pockel's coefficients of the stress-optic tensor; $\nu$ is the Poison's ratio and $\alpha$ is the coefficient of the thermal expansion of silica. For our specific situation where the SMF is a standard G.652 silica fiber with an optical carrier wavelength of 1550 nm at a room temperature, the factors in Eq. (S27) can be calculated, given by

$$\frac{\varepsilon_{n_{eff}}}{n_{eff}} = k_s \cdot \varepsilon_s + k_T \cdot \varepsilon_T \tag{S28}$$

where we have $k_s = 0.78 \times 10^{-6}\,\mu\varepsilon^{-1}$ and $k_T = 6.67 \times 10^{-6}\,°C^{-1}$ for silica fibers. The change of effective refractive index of SMF leads to a drift of eigenfrequency of the WSR. Substitute Eqs. (S1) and (S5) to Eq. (S28), we have

$$\begin{aligned}\varepsilon_{\omega_m^{(1)}} &= -\frac{2m\pi c}{n_{eff}^2 L}\varepsilon_{n_{eff}} \\ &= -\frac{2m\pi c}{n_{eff} L}\left(k_s \cdot \varepsilon_s + k_T \cdot \varepsilon_T\right)\end{aligned} \tag{S29}$$

where $L$ is the length of the SMF and $c$ is the light velocity in vacuum. In a laboratory environment, it is reasonable to assume disturbance magnitudes of $\varepsilon_s = 1\,\mu\varepsilon$ and $k_T = 0.1\,°C$. The corresponding eigenfrequency shifts are 7.8 kHz and 6.7 kHz, respectively, which are in the same order of magnitude compared to the FSR of the WSRs and thus will have nonnegligible effect on the operation of the PT symmetry. Specifically, for PT symmetry based on spatial duplex architecture (*2, 3, 6, 7*), temperature and vibration affect two spatial loops differently, and thus are differential mode perturbations. The perturbed Hamiltonian and eigenfrequencies are given by Eqs. (S23) and (S24), respectively. With $\varepsilon_{\omega_m^{(1)}}$ in the order of several kHz, the eigenfrequencies of supermodes experience strong bifurcation, which could lead to the breaking of PT symmetry. On the other hand, for the WSR-based PT-symmetric architecture, both PT-symmetric subspaces are enclosed in the same spatial loop. Temperature and vibration perturbations are applied to the two WSRs identically, and thus are common mode perturbations. Hence, the perturbed Hamiltonian and eigenfrequencies are given by Eqs. (S25) and (S26), respectively. No eigenfrequency bifurcation will be observed. Instead, the degenerate eigenfrequency shifts toward the same direction, indicating that the operation of PT symmetry is more stable in a WSR-based single spatial loop architecture.

7. Effect of temperature-dependent chromatic dispersion

In the above Section, we proved that WSR-based PT-symmetric system is immune to ambient temperature disturbance that is introduced via thermal expansion or the temperature-dependent refractive index. In our experiment, the eigenmode alignments between the two WSRs are achieve by tuning the wavelength spacing between the two optical carrier according to the dispersion coefficient of the SMF. Note that the dispersion of an SMF is temperature dependent, the temperature disturbance can also affect the PT symmetry operation by changing the dispersion. For the SMF with a fiber type of NZ-DSF, the thermal coefficient describing the relationship between chromatic dispersion and temperature is approximately given by (*8*)

$$\frac{1}{L} \cdot \frac{dD}{dT} = 0.0025 \text{ ps/nm/km/°C} \tag{S30}$$

In our experiment, the length of the SMF is 10 km and the magnitude of temperature disturbance is assumed to be 0.1 °C. We calculate the resulted chromatic dispersion perturbation to be $\varepsilon_D = 0.0025$ ps/nm. For carrier wavelength spacing of 2.353 nm that is chosen based on Eq. (S6), the change of dispersion will lead to a mismatch between the two WSRs. Again, substitute Eq. (S30) to Eqs. (S1) and (S5), we have

$$\frac{\varepsilon_{\omega_m^{(1)}}}{\varepsilon_D} = -\frac{2m\pi\Delta\lambda}{\left[\tau_0 + D \cdot \Delta\lambda\right]^2} = 3.08 \times 10^3 \text{ rad/s} \cdot \text{nm/ps} \tag{S31}$$

and the eigenfrequency perturbation $\varepsilon_{\omega_m^{(1)}}$ is 7.7 rad/s or 1.2 Hz, which is small and negligible compared to the FSR of the WSRs or the oscillating frequency. The temperature-dependent chromatic dispersion affects the two WSRs differently as the carriers are of different wavelengths, and thus is a differential mode perturbation. The perturbed Hamiltonian and the eigenfrequencies should have the form of Eqs. (S23) and (S24), respectively. Although eigenfrequency bifurcation is present in this situation, the magnitude of eigenfrequency perturbation $\varepsilon_{\omega_m^{(1)}}$ is only in the order of 1 Hz, which is more than three order of magnitude lower than the 7.8 and 6.7 kHz eigenfrequency perturbation that temperature and vibration can introduce to a spatial PT-symmetric system, i.e., a PT-symmetric system in the wavelength space can be over 1000 times more resilient to environmental perturbation compared to its spatial counterpart.

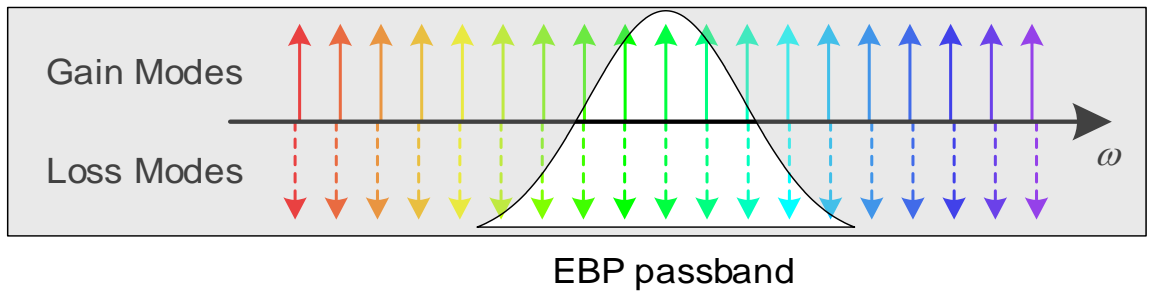

**Fig. S1.**
**Alignment of the eigenfrequencies of the gain and loss modes in the two WSRs.** The tuning of carrier wavelength in conjunction with the chromatic dispersion in the SMF are adopted to align the eigenmodes between the two resonators.

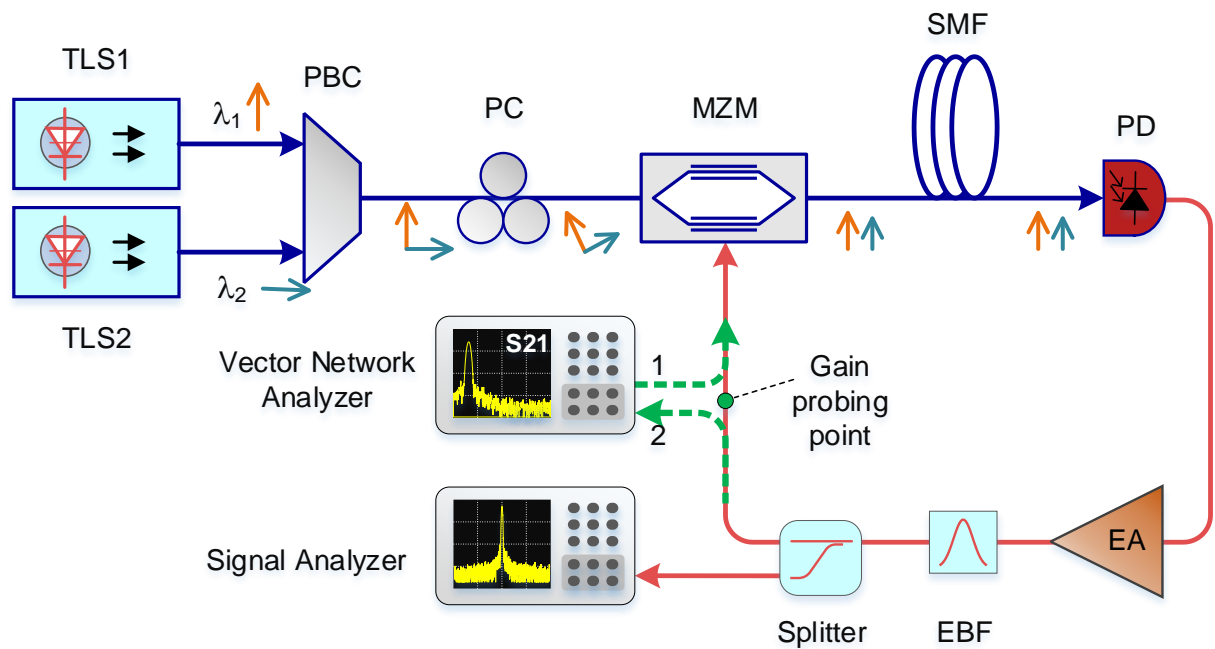

**Fig. S2.**

**Gain and loss coefficients probing and tuning in the WSRs.** A Vector Network Analyzer is deployed at the gain probing point between the microwave power splitter and the MZM. The measured loop gain coefficient, or S21 coefficient, is proportional to the square of the power of the optical carrier.

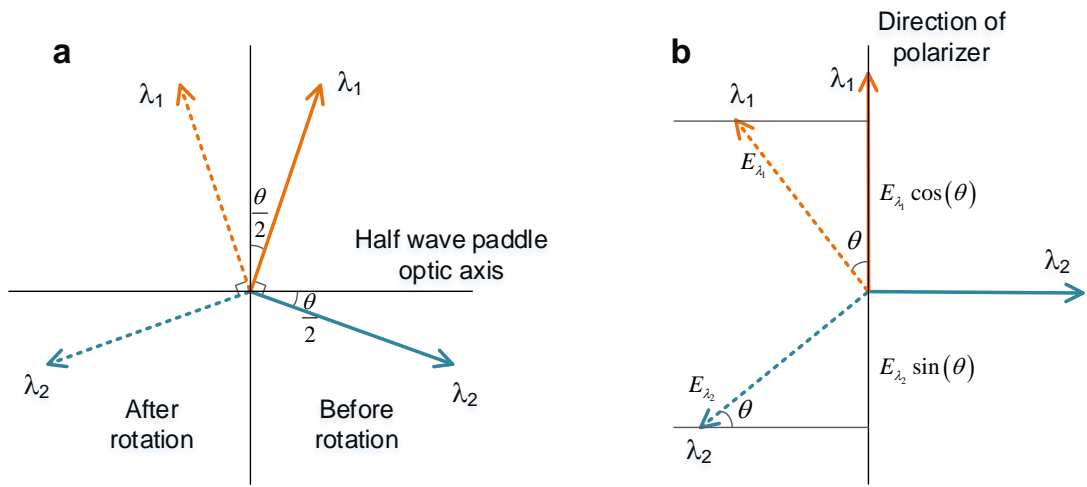

**Fig. S3.**

**Tuning PC changes carrier power and is used to fine-tune the gain and loss coefficients in the two WSRs. a**. Polarization rotation with the PC; **b**. carrier power tuning after propagating through a polarizer and polarization rotation using the PC.

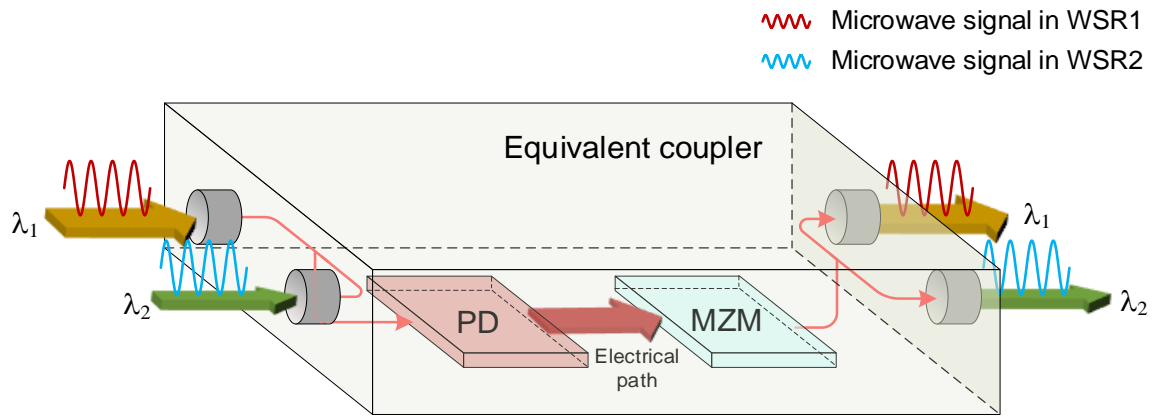

**Fig. S4.**
**The electrical path in the WSRs operating equivalently to a microwave coupler.** The microwave signal is combined and distributed indistinctively from and to the two carriers. The electrical path can be modelled as a coupler with a coupling ratio of 1:1, and the power of optical carrier is interpreted as the microwave insertion loss in the optical path in the WSRs.

**Movie S1.**

The electrical spectrum of the generated microwave signal from the wavelength-space PT-symmetric OEO. Single-mode oscillation is achieved with high stability.